\documentclass[twocolumn]{aastex63}

\usepackage{amsmath}
\usepackage{amssymb}
\usepackage{amsthm}
\usepackage{natbib,aasdefs,url,bm}
\usepackage{graphicx}
\usepackage{color}
\usepackage{mathtools,  epsfig, lipsum}
\usepackage{subfigure}
\usepackage{textcomp}

\usepackage{multirow,booktabs,colortbl,tabularx}
\usepackage{enumitem}
\usepackage{booktabs}
\usepackage{hyperref}

\shorttitle{A cross correlation of IceCube neutrinos and galaxies}
\shortauthors{Fang, Banerjee, Charles, \& Omori}

\begin{document}

\title{A Cross-Correlation Study of High-energy Neutrinos and Tracers of  Large-Scale Structure}

\author{Ke Fang} 
\affil{NHFP Einstein Fellow}
\affil{Kavli Institute for Particle Astrophysics and Cosmology (KIPAC), Stanford University, Stanford, CA 94305, USA}

\author{Arka Banerjee} 
\affil{Kavli Institute for Particle Astrophysics and Cosmology (KIPAC), Stanford University, Stanford, CA 94305, USA}
\affil{SLAC National Accelerator Laboratory, 2575 Sand Hill Road, Menlo Park, CA 94025, USA}

\author{Eric Charles}  
\affil{Kavli Institute for Particle Astrophysics and Cosmology (KIPAC), Stanford University, Stanford, CA 94305, USA}
\affil{SLAC National Accelerator Laboratory, 2575 Sand Hill Road, Menlo Park, CA 94025, USA}

\author{Yuuki Omori}
\affil{Kavli Institute for Particle Astrophysics and Cosmology (KIPAC), Stanford University, Stanford, CA 94305, USA}

\begin{abstract}
The origin of the bulk of the astrophysical neutrinos detected by the IceCube Observatory remains a mystery.   Previous source-finding analyses compare the directions of IceCube events and individual sources in astrophysical catalogs.  The source association method is technically challenging when the number of source candidates is much larger than the number of the observed astrophysical neutrinos. We show that in this large source number regime,  a cross-correlation analysis of neutrino data and source catalog can instead be used to constrain potential source populations for the high-energy astrophysical neutrinos, and provide spatial evidence for the existence of astrophysical neutrinos. We present an analysis of the cross-correlation of the IceCube 2010-2012 point-source data and a WISE-2MASS galaxy sample. While we find no significant detection of cross-correlation with the publicly available neutrino dataset, we show that, when applied to the full IceCube data, which has a longer observation time and higher astrophysical neutrino purity, our method has sufficient statistical power to detect a cross-correlation signal if the neutrino sources trace the Large Scale Structure of the Universe.
\end{abstract}

\keywords{high-energy neutrinos, large-scale structure}

\section{Introduction}

The existence of an astrophysical neutrino population above $\sim 100$~TeV has been established by the IceCube Observatory \citep{1242856, 2016ApJ...833....3A, 2019ICRC...36.1004S,  2019ICRC...36.1017S}. The discovery is based on an excess of the observed flux over the atmospheric background in several detection channels. No signatures of clustering of astrophysical neutrinos has been measured. 
The origin of these cosmic neutrinos remains unknown. Point-source searches have been carried out, in the form of blind searches that scan the sky with sub-degree grids \citep{2018arXiv181107979I, 2019ApJ...886...12A, 2019arXiv191008488I}, and source association searches that stack the likelihoods of sources in a given catalog (e.g., \citealp{2017ApJ...835...45A, 2019arXiv191111809I}).  None of the searches has led to a robust identification of the sources of the bulk of the observed neutrinos.

Formally, the likelihood stacking analysis requires computation of the probability of each neutrino coming from every source in the galaxy catalog. The complexity of such an analysis scales with the number of sources in a catalog, and can be computationally challenging due to factors such as spatial correlations of closely spaced sources. 
However, high-energy neutrinos may plausibly come from sources from a large population with apparent number density $\gtrsim 10^{-6}\,\rm Mpc^{-3}$ (more than a million sources within redshift $z\sim 1$), such as star-forming galaxies (e.g., \citealp{2014JCAP...09..043T, 2014PhRvD..90j3005F}) and galaxy clusters (e.g., \citealp{2018NatPh..14..396F}). In this paper, we present a novel approach to test the connection of IceCube events with potential sources that are large in population.     

The two-point correlation function is a common statistical tool used to describe the distribution of galaxies \citep{1980lssu.book.....P}. It is a measure of the excess probability of finding a pair of data points at some separation $R$ compared to a random distribution of points. The cross-correlation generalizes this definition to the case when the two data points are drawn from different datasets. In this work, we show that the cross-correlation function, or its equivalent in harmonic space, the cross power spectrum, can be applied to the study of high-energy neutrinos. 
We have developed the framework and an analysis pipeline to compute the cross-correlation between IceCube events and galaxy samples and applied this methodology to the IceCube 2010--2012 point-source dataset and a galaxy catalog based on the Wide-Field Infrared Survey Explorer (\citealp{2010AJ....140.1868W}, WISE) and the 2-Micron All-Sky Survey (\citealp{2006AJ....131.1163S}, 2MASS) infrared databases.  Unlike the stacking analyses that were applied to test, e.g., finding whether neutrinos come from sources in the Fermi-2LAC Blazar catalog \citep{2017ApJ...835...45A},
the cross-correlation analysis outlined here does not scale in computational cost with the number of sources, and does not require the use of a catalog of the exact neutrino source class, as long as neutrino sources and the galaxy sample used in the analysis both trace the same underlying large-scale density field. Detection of cross-correlations with different galaxy samples covering different galaxy types and redshift distributions could potentially narrow down the source classes and redshift range of origin for the high-energy neutrinos.

The paper is organized as follows. The method is laid out in Section~\ref{sec:method} and Appendix~\ref{appendix:mixCl}. The analysis is described in Section~\ref{sec:analysis} and Appendices~\ref{appendix:pointSourceData}, \ref{appendix:syntheticData}, \ref{appendix:GalaxySample}. Findings are summarized in Section~\ref{sec:results} and discussion and conclusions are presented in Section~\ref{sec:discussion}.

\section{Cross-correlation of neutrinos and galaxies}\label{sec:method}
Let $n_g(\mathbf x)$ and $n_\nu(\mathbf x)$ represent the number density of galaxies and the number density of neutrino events detected by IceCube at some position $\mathbf x$ on the plane of the sky respectively. 
For both fields, we can define the overdensity field as 
\begin{equation}
\delta (\mathbf x) \equiv \frac{n(\mathbf x) - \bar{n}}{\bar{n}} \,,
\end{equation}
where $\bar{n}$ is the average value of $n(\mathbf x)$ over all directions used in the analysis.  Below, we refer to  $\delta$ as the fluctuation or overdensity field interchangably. 

The two-point cross correlation function of the galaxy and neutrino fluctuation fields, and $C^{g\nu}$, is defined as  

\begin{equation}
C^{g\nu}_\ell = \frac{4\pi}{2\ell + 1}\,\int d\cos\theta\, \langle\delta_g(\mathbf x)\delta_\nu(\mathbf x')\rangle \,P_\ell^*(\cos\theta)\,,
\label{eq:cl_def}
\end{equation}

where $\theta$ is the angle between the two directions $\mathbf x$ and $\mathbf x'$, $P_\ell$  is the Legendre polynomial, and the average is performed in the $\mathbf x$ space. 
The term $C^{g\nu}_\ell$ denotes the contribution of the harmonic $\ell$ to the correlation function, 
\begin{equation}\label{eqn:Cl}
C^{g\nu}_\ell = \frac{1}{f_{\rm sky}(2\ell+1)}\sum_{m=-\ell}^{\ell} a^{g*}_{\ell m}\,a^\nu_{\ell m}
\end{equation}
where $a_{\ell m}$ are the coefficients to decompose the fluctuation field into spherical harmonics, $\delta (\theta,\phi) = \sum_{l=0}^{\infty}\sum_{m=-\ell}^{\ell}a_{\ell m}\,Y_{\rm \ell m}(\theta,\,\phi)$. $f_{\rm sky}$ is the fraction of the sky from which neutrino and galaxy data is used for analysis. 

Assuming that extremely high-energy neutrinos are produced in some specific  source population(s) --- \textit{e.g.} blue star-forming galaxies --- \textit{which trace the same underlying matter density modes as the galaxies used in the analysis}, then on large scales ($\ell\lesssim 300$) we can write \citep{1986ApJ...304...15B}
\begin{align}
    C_\ell^{gg} &= b_g^2 C_\ell^{mm} \nonumber \\
    C_\ell^{ss} &= b_s^2 C_\ell^{mm} \nonumber \\
    C_\ell^{gs} &= b_g b_s C_\ell^{mm} \,,
\label{eq:bias_def}
\end{align}
where $b_g$, $b_s$ are the bias parameters of the two populations\footnote{The bias parameter relates the clustering of the peaks of a Gaussian random field to the clustering of the underlying field \citep{1984ApJ...284L...9K}. More massive halos generally form on rarer peaks of the initial field, and have a higher bias parameter. $b_{s}$ is the bias parameter of the galaxies that source the neutrinos}, and $C_\ell^{mm}$ represents the power spectrum of the underlying density field. If the astrophysical neutrinos detected by IceCube represent a Poisson sampling from the source population, then the cross correlation spectrum of these events with the galaxy catalog can be written as
\begin{equation}
C_\ell^{g\,\rm astro} = f_{\rm g}\,C_\ell^{gg} \,.
\label{eq:fg_def}
\end{equation}
Comparing Equation~\ref{eq:fg_def} to Equation~\ref{eq:bias_def}, we see that $f_g = b_s/b_g$.
The value of $f_{\rm g}$ depends on value of the bias of both the galaxy sample used ($b_g$), as well as the source population ($b_s$). However, for most galaxy samples, and for most of the proposed source populations, the bias values are $\sim \mathcal O(1)$, and therefore, $f_g$ is also expected to be $\mathcal O(1)$. For example, if we use BOSS-CMASS galaxies around $z\sim 0.5$ as the galaxy sample, $b_g\sim 2$ \citep{2013MNRAS.432..743N}, and if the source population is blue star-forming galaxies in a similar redshift range, $b_s\sim 1$ (e.g., \citealp{2017A&A...599A..62L}), implying $f_g \sim 0.5$. 

It should be noted that the IceCube data is dominated by atmospheric neutrinos up to $\sim 100$~TeV in the northern sky, and by cosmic rays up to the highest energies in the southern sky. For a mixed population of astrophysical and atmospheric neutrinos from cosmic-ray interactions, the cross power spectrum satisfies
\begin{equation}
C_{i,\ell}^{g\nu} = f_{\rm astro, i}\,C_{i,\ell}^{g\,\rm astro} + (1-f_{\rm astro, i})\,C_{i,\ell}^{g\,\rm atm}. 
\end{equation}
with $f_{\rm astro, i}$ being the fraction of astrophysical events in energy bin $i$ (see Appendix~\ref{appendix:mixCl}). 

Assuming that the energy bins are independent \citep{Aartsen_2014}, we define the likelihood function by 
\begin{equation}\label{eqn:logL}
\log\,{\cal L} ({\mathbf C}_\ell^{g\nu} |{\mathbf f}_{\rm astro})= -  \sum_{i, \ell}\,\frac{\left(C_{i, \ell}^{g\nu} - \langle C_{i, \ell}^{g\nu}(f_{\rm astro, i})\rangle\right)^2}{2\,\left(\sigma_{i,\ell}^{g\nu}\right)^2}. 
\end{equation}
As atmospheric neutrinos and muons do not trace the distribution of galaxies, $\langle C_{i, \ell}^{g\,\rm atm}\rangle=0$, and the expected mean cross-correlation of a combined sample of astrophysical and atmosphere neutrinos is $\langle C_{i, \ell}^{g\nu}\rangle = f_g\,f_{\rm astro, i}\,C_\ell^{gg}$. The expected standard deviation, $\sigma_{i,\ell}^{g\nu}$, of a combined sample is obtained by running a set of  Monte Carlo simulations that contain both astrophysical and atmospheric events and taking the standard deviation of the cross-correlation values obtained in each sample (see Appendix~\ref{appendix:syntheticData}). In general we find that $\sigma_{i,\ell}^{g\nu}$ is insensitive to $f_{\rm astro, i}$, and is inversely proportional to the square root of the sample size $\propto N_\nu^{-1/2}$. 

The significance of a signal against the null hypothesis, defined as zero correlation between the neutrino sample and the source catalog, can be quantified by a test statistic, 
\begin{equation}
   {\rm TS} \equiv 2 \left[ \log {\cal L} (\hat{f}_g,\,\hat{\mathbf f}_{\rm astro})- \log {\cal L}(\mathbf{0})\right] \, ,
\end{equation}
where $\hat{f}_g$ and $\hat{\mathbf f}_{\rm astro}$ are the maximum likelihood values of $f_g$ and $\mathbf{f}_{\rm astro}$, respectively. 
If the neutrino and galaxy fluctuation fields are Gaussian, the TS should follow the chi-square distribution with degrees of freedom equal to the number of energy bins used for the likelihood evaluation \citep{wilks1938}.

\begin{figure*}
\includegraphics[width=0.43 \linewidth] {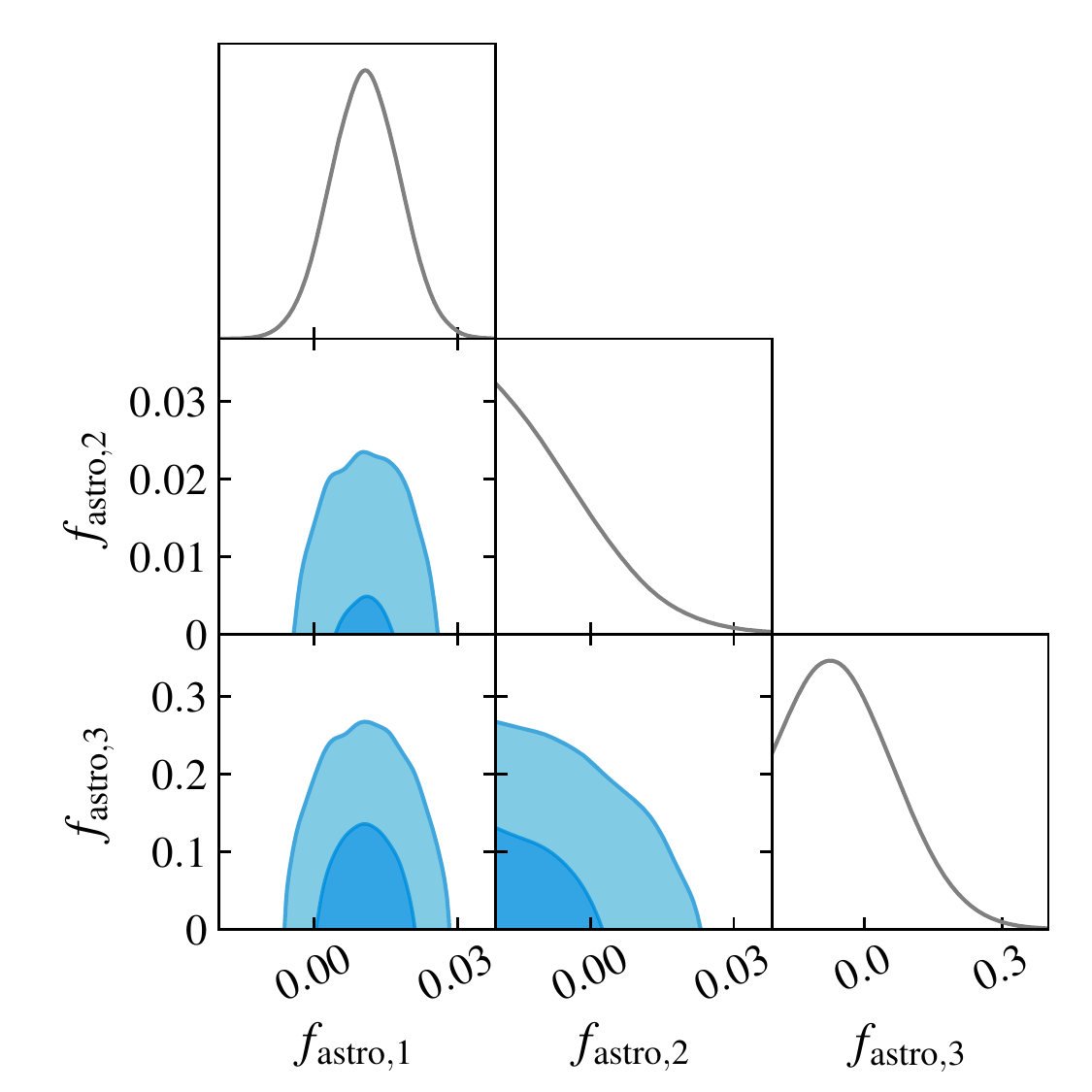} 
\includegraphics[width= 0.57 \linewidth] {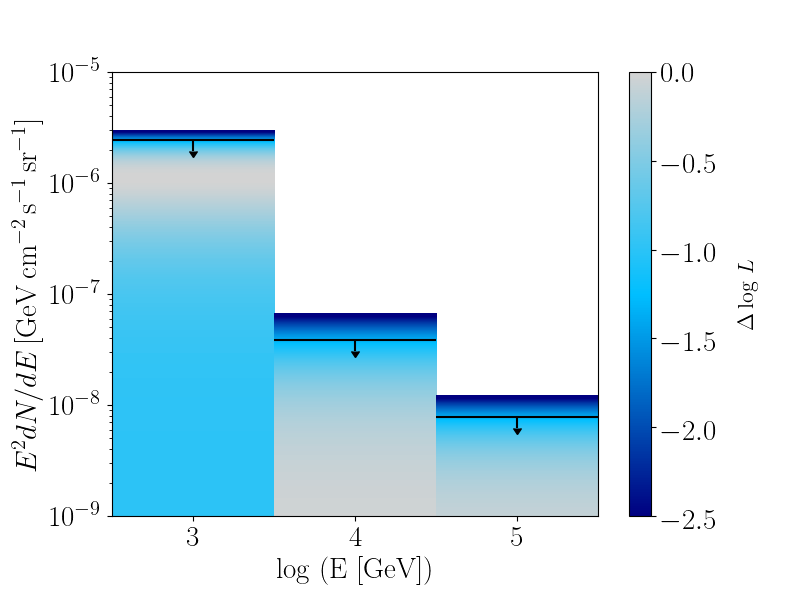}  
\caption{\label{fig:IC3yr}
Left: the 1-dimensional posterior distributions on $f_{g,\rm astro,i}$ and 2-dimensional marginalized contours of the model.   
The contours indicate the 68\%, 95\%, and 99\% C.L. regions for individual parameters found by a MCMC sampling of the parameter space. Right:  upper limits at 95\% C.L. on the energy flux of astrophysical muon neutrinos in the public three-year point-source data whose overdensity field cross correlates with the overdensity field of the  WISE-2MASS galaxy sample. The color indicates the distance in log-likelihood, for given energy flux, from the peak of the log-likelihood found by maximizing equation~\ref{eqn:logL}. The conversion between $\mathbf{f}_{\rm astro}$ (assuming $f_g=1$) and energy flux is described by equation~\ref{eqn:flux}.
}
\end{figure*}

\section{Analysis setup}\label{sec:analysis}

An ideal neutrino dataset for the cross correlation study would have full-sky coverage, good angular resolution, large sample size and high purity of astrophysical neutrino events. The veto techniques of IceCube 
 (e.g.,  \citealp{1242856, PhysRevD.91.022001}) reduce cosmic-ray background by selecting neutrino events that interact within the detector boundary. The through-going tracks from the northern hemisphere also provide a clean neutrino sample, as up-going muons are suppressed by the earth \citep{2019ICRC...36.1017S}.  These contained-vertex and through-going track event samples are suitable for the proposed analysis. 
 
 Of the public datasets\footnote{https://icecube.wisc.edu/science/data/access}, only the point-source dataset\footnote{https://icecube.wisc.edu/science/data/PS-3years}\citep{Aartsen_2017} includes both the full direction information and the corresponding effective area tables. Therefore, we tailored the public point-source dataset for a demonstration of the cross correlation analysis.  

The point-source data is composed of track-like events with angular resolutions ranging from $\sim 1^\circ$ around 1~TeV to  $\sim0.4^\circ$ above 10~TeV \citep{Aartsen_2017}. We bin the data into HEALPix\footnote{http://healpix.sf.net} sky-maps with $N_{\rm side}=128$ using the \textsc {healpy} package\footnote{https://healpy.readthedocs.io/en/latest/}\citep{2005ApJ...622..759G,Zonca2019}. To ensure that enough counts are available for the analysis, we group events into decade-wide logarithmic energy bins ranging from $10^{1.5}$ to $10^{8.5}$~GeV. To avoid the large muon cosmic-ray background, we only use events from the northern sky, defined as declination angle Dec~$> -5^\circ$ \citep{Aartsen_2017}. The counts map, distribution of the zenith angle, and the auto-correlation of the point-source data are presented in Appendix~\ref{appendix:pointSourceData}. 

As shown in Figure~\ref{fig:IceCubeAuto}, the spatial distribution of the effective area of the IceCube point-source data is smoother than that of three-year neutrino events on angular scales $\ell \gtrsim 50$.  Event counts thus trace the source distribution on these small angular scales.  For this study we prefer counts to flux as an indicator of the source distribution, as 
single events in spatial bins with very low exposure can result in anomalously large fluxes. 

We generate ``negative control'' samples of synthetic atmospheric neutrino data based on the zenith angle distribution of observed events (see Appendix~\ref{appendix:syntheticData}). To test the sensitivity of our method in finding a cross-correlation signal, we sample astrophysical events from the density field of our galaxy sample (see below) and set  $\mathbf{f}_{\rm astro}$ based on the diffuse muon neutrino flux in the IceCube ten-year data (Figure~3 of \citealp{2019ICRC...36.1017S}), ${\mathbf f}_{\rm astro}= {\mathbf f}_{\rm astro}^{\nu_\mu}$. The sample size of the synthetic data is randomly generated from a Poisson distribution centered at the observed event count. As the IC79-2010 data 
has different spatial and energy distributions from the IC86-2010 and 2012 data, we generate synthetic data year by year and sum the resulting count maps.

The all-sky galaxy catalog used in this work is constructed following \citet{2015MNRAS.448.1305K} to  combine photometric information of the WISE \citep{2010AJ....140.1868W} and 2MASS \citep{2006AJ....131.1163S} infrared databases 
(see Appendix~\ref{appendix:GalaxySample} for more details). The region with Galactic latitude $|b|<10^\circ$ is masked to avoid Galactic foregrounds.  The median redshift of the sample is $z\approx 0.14$.

The $a_{\ell m}$ coefficients of the neutrino and galaxy over-densities, and the $f_{\rm sky}$ accounting for neutrino and galaxy masks are used to compute $C_\ell^{g\nu}$ following equation~\ref{eqn:Cl}.  We set $\ell_{\rm min} = 50$ when computing the likelihood in Equation~\ref{eqn:logL} to avoid effects from the non-uniform IceCube exposure and the masks of neutrino and galaxy data (see Appendix~\ref{appendix:pointSourceData}). The results, however, do not significantly depend on $\ell_{\rm min}$ as long as $\ell_{\rm min}>$ a few.  
The standard deviations of the model, $\sigma_{i,\ell}^{g\nu}$, are precomputed from 500 realizations of synthetic data. We use three energy bins $i=1, \,2, \,3$ (uniform in the logarithm from 0.3~TeV to 300~TeV) for the likelihood calculation, considering that the data is heavily dominated by atmospheric events below $\sim$3~TeV, and that the three-year point-source data has no events above 3~PeV in the northern sky.

The fractions $f^i_{\rm astro}$ in the different energy bins $i$ are assumed to be independent, and are coupled with $f_g$. The value of $f_g$ depends on the redshift of the galaxy sample and the type of neutrino source, but is generally on the order of unity (see Section~\ref{sec:method}). We thus set $\mathbf{f}_{g,\rm astro} \equiv f_g\,\mathbf{f}_{\rm astro}$ and allow a higher upper bound for $|f_{g,\rm astro, i}|$ to account for the cases where $f_g > 1$. We also allow negative $f_{g,\rm astro, i}$ to describe anti-correlation of neutrino sources and galaxy catalogs. The results are insensitive to the upper bound, and we obtain similar results with upper bounds ranging from 1 to 4.

\begin{figure}[ht]
\includegraphics[width= \linewidth] {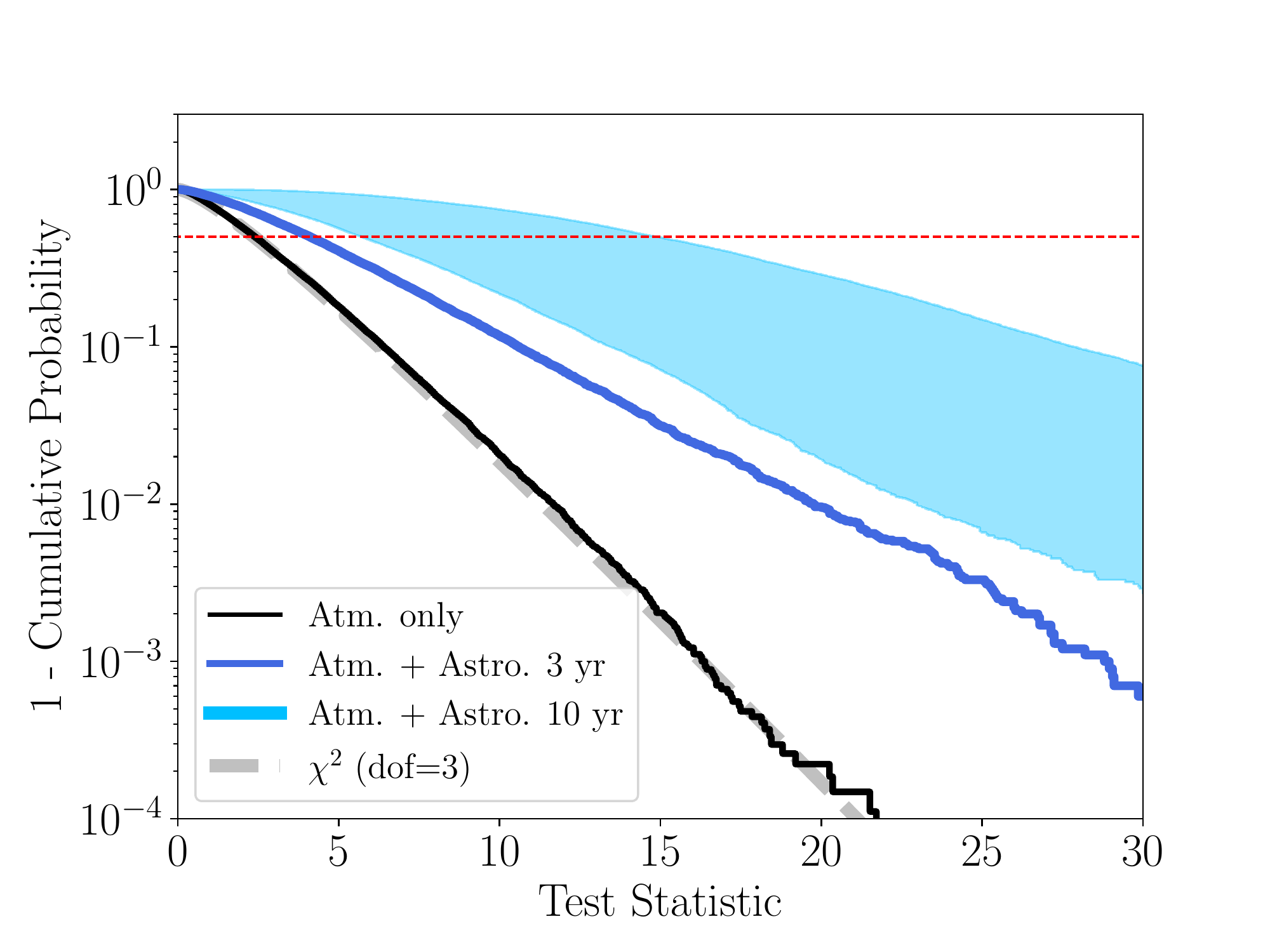}  
\caption{\label{fig:TS_distribution}
Cumulative distribution of test statistic of the cross correlation of synthetic neutrino data and the WISE-2MASS galaxy sample. The black curve indicates neutrino samples composed of only atmospheric events from ten years of observation. For comparison, the grey dashed line shows the probability function of a chi-square distribution with three degrees of freedom. The dark blue curve and light blue filled region indicate neutrino samples that contain astrophysical neutrinos from galaxy-like sources with three and ten years of observation respectively. The bounds of the filled region correspond to different input values of $\mathbf{f}_{g,\rm astro}$. See Section~\ref{sec:results} for details. The red dashed line denotes the 50\% cumulative probability. With a contained event sample such as HESE and ten-year observation (similar to the scenario represented by the upper edge of the light blue region), the optimistic scenario discussed in the text would have a 50\% chance of yielding a cross correlation test statistic of 15 or greater, corresponding to a detection significance of $3.1\sigma$.
}
\end{figure}

\section{Results}\label{sec:results}


Cross-correlating the northern-sky events in the IceCube three-year point-source data and the WISE-2MASS galaxy sample leads to best-fit $\hat{f}_{g,\rm astro, 1, 2, 3} = (0.011,  -0.027, -0.076)$ and ${\rm TS} = 4.3$ (corresponding to $1.2\,\sigma$ confidence level in a two-tail normal distribution). We find no evidence of astrophysical neutrinos in the tailored dataset that follow the spatial distribution of the WISE-2MASS galaxy sample. 

We take a Bayesian approach to sample the parameter space of $\mathbf{f}_{g,\rm astro}$ using Markov chain Monte Carlo (MCMC) via emcee\footnote{https://emcee.readthedocs.io/en/stable/}.  We adopt a uniform prior probability for $-4<f_{g,\rm astro, i}<4$. The left panel of Figure~\ref{fig:IC3yr} presents the findings of a ensemble sampler with 640~walkers and 500~steps. The black contours indicate the 68\%, 95\%, and 99\% confidence intervals for $f_{g,\rm astro, 1, 2, 3}$ accordingly. 

The fraction of astrophysical events that cross correlate with galaxies can be converted into an energy flux through 
\begin{equation}\label{eqn:flux}
    \phi_\nu (E_i) \approx \frac{ N_{\nu, i}\,f_{\rm astro, i} }{ \Delta E_i\,\Delta t\,f_{\rm sky}\,4\,\pi\,\bar{A}_{\rm eff, i}},
\end{equation}
where $N_\nu, i$ is the number of neutrino events in bin $i$, $\Delta t$ is the active observation time of IceCube and $\bar{A}_{\rm eff, i}$ is the mean of the weighted effective area (as defined in equation~\ref{eqn:weightedAeff}) over the unmasked region.  To avoid negative flux, we assume $f_g=1$ and use a one-sided test $\mathbf{f_{\rm astro}} > 0$ to obtain the fractions before converting them to flux. Depending on the significance of the cross correlation, we quote $\phi_\nu$ as an upper limit at the 95\% C. L. (when TS~$< 4$) or a data point with $1\,\sigma$ error bars (when TS~$>4$, corresponding to $2\sigma$ with one degree of freedom). The upper limits of $\mathbf{f}_{\rm astro}$ at 95\% C.L. are $(0.022, 0.016, 0.16)$. The right-hand panel of Figure~\ref{fig:IC3yr} presents the converted upper limits to the energy flux of astrophysical neutrinos whose sources are distributed like the WISE-2MASS galaxy sample. 

To demonstrate the efficiency of the cross-correlation method, Figure~\ref{fig:TS_distribution} presents the cumulative probability distribution of the TS of cross correlation between the WISE-2MASS galaxy sample and the synthetic three-year and ten-year data. The plot is made from $10^4$ realizations of synthetic data that  contain only atmospheric neutrinos or atmospheric neutrinos plus astrophysical neutrinos sampled from the galaxy density field. The TS distributions of the background-only samples agree with the chi-square distribution, confirming the Gaussianity of our likelihood function. The filled regions denote the TS distributions of mixed-population neutrino samples from three-year (light blue) and ten-year (dark blue) observations. The lower bounds of the filled regions correspond to a muon neutrino-like sample, with input ${\mathbf f}_{\rm astro}= {\mathbf f}_{\rm astro}^{\nu_\mu}$ suggested by the diffuse muon neutrino analysis (\citealp{2019ICRC...36.1017S}; also see Appendix~\ref{appendix:syntheticData}). The upper bounds correspond to a more optimistic scenario with ${\mathbf f}_{\rm astro}= 2 \,{\mathbf f}_{\rm astro}^{\nu_\mu}$ motivated by the fraction of astrophysical events above 30~TeV in the High-energy starting events (HESE) sample \citep{2019ICRC...36.1004S}. With ten years of full-sky IceCube data and efficient selection of astrophysical events (corresponding to optimistic $\mathbf{f}_{\rm astro}$), the technique presented here should be able to detect cross-correlations between the astrophysical neutrino events seen by IceCube and different tracers of Large-Scale Structure, thereby allowing us to constrain the source populations generating the high-energy neutrinos.

\section{Discussion and Conclusions}\label{sec:discussion}
High-energy neutrinos are a unique messenger of hadronic processes of the Universe at extreme energies. Understanding their origin is a crucial task in Neutrino Astronomy. Previous source searches have focused on the association of IceCube events with individual sources in a catalog, and are thus limited to source classes with relatively small populations. In this paper, we have performed the first cross-correlation analysis between the IceCube events and a tracer of the Large-Scale Structure -- galaxies from a WISE-2MASS catalog. A non-zero cross-correlation is expected if the source population generating high-energy neutrinos, such as star-forming galaxies and galaxy clusters, trace the same underlying matter density modes as the galaxy sample. Such a detection would be the first spatial evidence for the astrophysical origin of the high-energy neutrino events. Further, in the scenario of complete galaxy samples and sufficient astrophysical neutrino events at IceCube, detections and non-detections of cross-correlation with each galaxy sample can be used to narrow down the sources of the IceCube neutrinos. 

Using the the publicly available three-year point-source dataset from IceCube we do not find a significant detection of cross-correlation with the WISE-2MASS galaxy sample. However, we show that if the analysis is performed with the full IceCube data of contained events and northern-sky muon neutrino events, we should have a statistically significant detection of any true cross-correlation. We urge an immediate analysis followup by the high-energy neutrino experiments.


Apart from increasing the sample of neutrino events, improvements are also possible by optimizing the Large-Scale Structure tracer used in the analysis.
The cross-correlation analysis is best performed using a complete and clean tracer of the LSS. The current analysis uses a sample of galaxies constructed from the WISE and 2MASS infrared surveys. The sample has low stellar contamination and high completeness, but only contains nearby galaxies with a median redshift $z\approx 0.14$. To better search for cross-correlation with LSS, samples with higher number densities, different redshift ranges, and larger sky coverage can be used. For example, the Sloan Digital Sky Survey provides a wide-field coverage of galaxies out to $z\sim 0.8$ in the optical wavelength range \citep{2019arXiv191202905A}. It is also possible to use tracers of Large-Scale Structure other than galaxies. For example, the cosmic infrared background map from Planck \citep{2019ApJ...883...75L} consists of infrared emission from dusty galaxies and traces the star formation history. Measurements of the weak lensing shear from the Dark Energy Survey \citep{2018MNRAS.475.3165C} can also be used as a proxy for the underlying matter field on large scales. 

As pointed out in Section \ref{sec:method},   $\mathbf{f}_{\rm astro}$ and $f_g$ are currently strongly correlated. However, we note that this degeneracy may be broken if the energy spectrum of the neutrino sample is known. Measurements of the diffuse astrophysical neutrino flux provide an estimation of $\mathbf{f}_{\rm astro}$ based on the modeling of the atmospheric neutrino contribution. Then the cross correlation analysis can derive $f_g$, which would inform the relation of the neutrino sources to the test sources in use. 

The analysis code we developed to perform cross correlation of IceCube events and galaxy samples is available\footnote{\hyperlink{https://github.com/KIPAC/nuXgal}{https://github.com/KIPAC/nuXgal}}. It is written in {\it Python} and uses {\it healpy} to perform calculations of spherical harmonics.


We thank Erik Blaufuss and Mike Richman for their helpful discussion about the IceCube effective area of the public point-source data. We thank Seth Digel for his extremely useful feedback on our manuscript.

\bibliographystyle{aasjournal}

\appendix
\restartappendixnumbering

\section{Cross correlations in the presence of a mixture of populations}\label{appendix:mixCl}

The neutrinos detected by IceCube in any given energy bin come from two distinct populations - atmospheric neutrinos and astrophysical neutrinos. The total number density of neutrinos detected at some point $n_\nu(\mathbf x)$ on the sky is given by
\begin{equation}
    n_\nu(\mathbf x) = n_{\rm astro }(\mathbf x)+ n_{\rm atm}(\mathbf x)\,.
\end{equation}
We can recast the equation above in terms of mean densities and overdensities:
\begin{equation}
    \bar n_\nu(1+\delta_\nu(\mathbf x)) =  \bar n_{\rm astro}\left(1+\delta_{\rm astro}(\mathbf x)\right)+ \bar n_{\rm atm}\left(1+\delta_{\rm atm}(\mathbf x)\right) \,. 
\end{equation}
By definition $\bar n_\nu = \bar n_{\rm astro}+\bar n_{\rm atm}$, and so,
\begin{equation}
    \delta_\nu (\mathbf x) =  f_{\rm astro} \delta_{\rm astro}(\mathbf x)+(1-f_{\rm astro})\delta_{\rm atm}(\mathbf x) \,,
\end{equation}
where $f_{\rm astro} = \bar n_{\rm astro}/\bar n_\nu$. 
Using Eq.~\ref{eq:cl_def}, we have 
\begin{align}
    \langle\delta_g(\mathbf x)\delta_\nu(\mathbf x')\rangle &= \sum_\ell\,\frac{2\,\ell + 1}{4\,\pi}\,C^{g\nu}_\ell\,P_\ell(\cos\theta) \nonumber \\
    \implies \left\langle\delta_g(\mathbf x)\left(f_{\rm astro} \delta_{\rm astro}(\mathbf x')+(1-f_{\rm astro})\delta_{\rm atm}(\mathbf x')\right)\right \rangle &= \sum_\ell\,\frac{2\,\ell + 1}{4\,\pi}\,\bigg(f_{\rm astro}C^{g \,{\rm astro}}_\ell + (1-f_{\rm astro})C^{g \,{\rm atm}}_\ell\bigg)\,P_\ell(\cos\theta) \,.
\end{align}
Comparing coefficients of $P_\ell(\cos\theta)$, we see that for a mixed population of astrophysical and atmospheric neutrino events with $f_{\rm astro}$ fraction of astrophysical neutrinos, the cross correlation with the galaxy sample can be written in terms of the individual cross correlations as 
\begin{equation}
    C^{g\nu}_\ell = f_{\rm astro}C^{g \,{\rm astro}}_\ell + (1-f_{\rm astro})C^{g \,{\rm atm}}_\ell
\end{equation}

\section{IceCube data analysis and synthetic data generation}

\subsection{IceCube All-sky Point-Source Data in 2010-2012}\label{appendix:pointSourceData}

\begin{figure}
\includegraphics[width= 0.49 \linewidth] {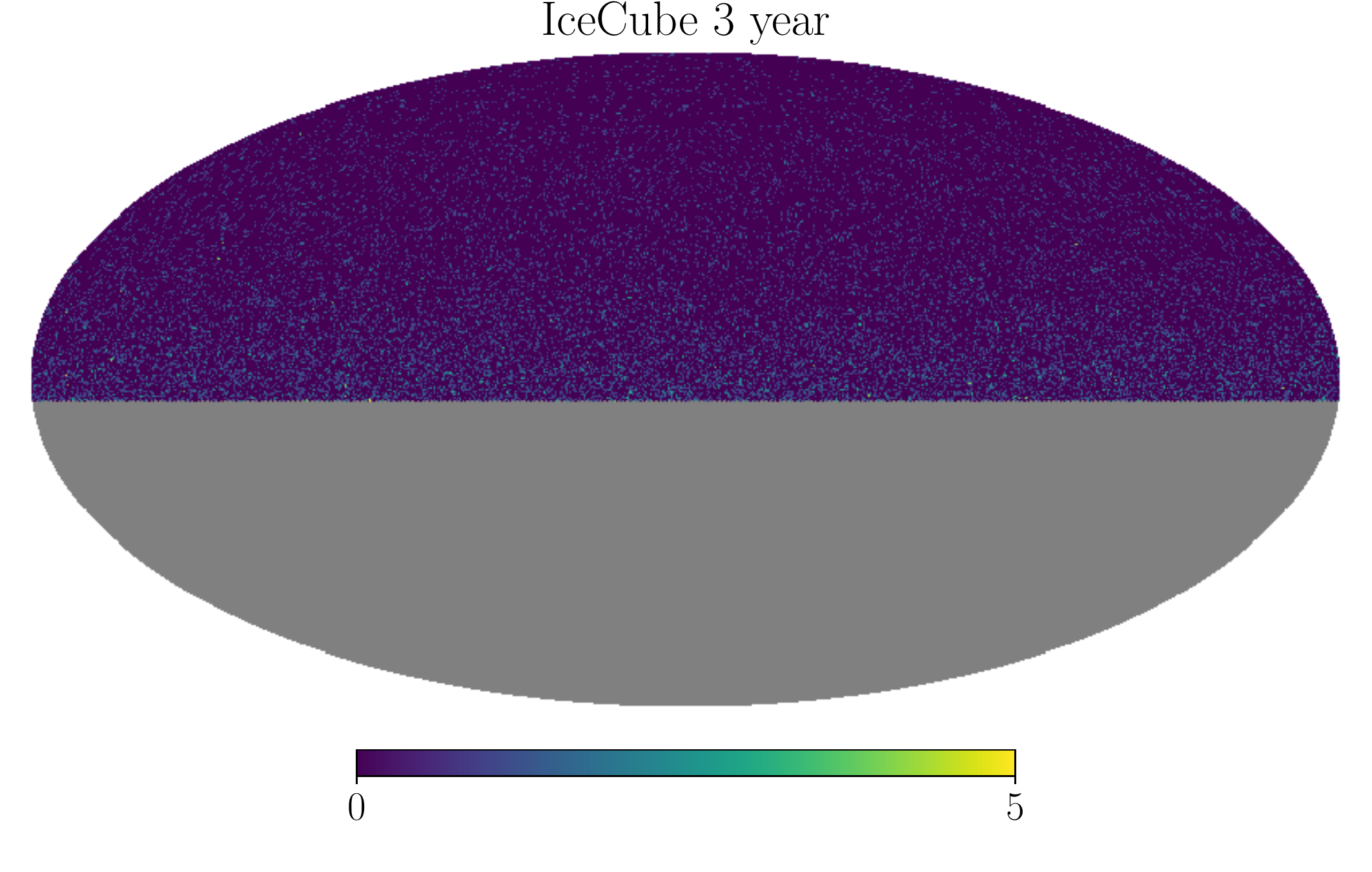}  
\includegraphics[width= 0.49 \linewidth] {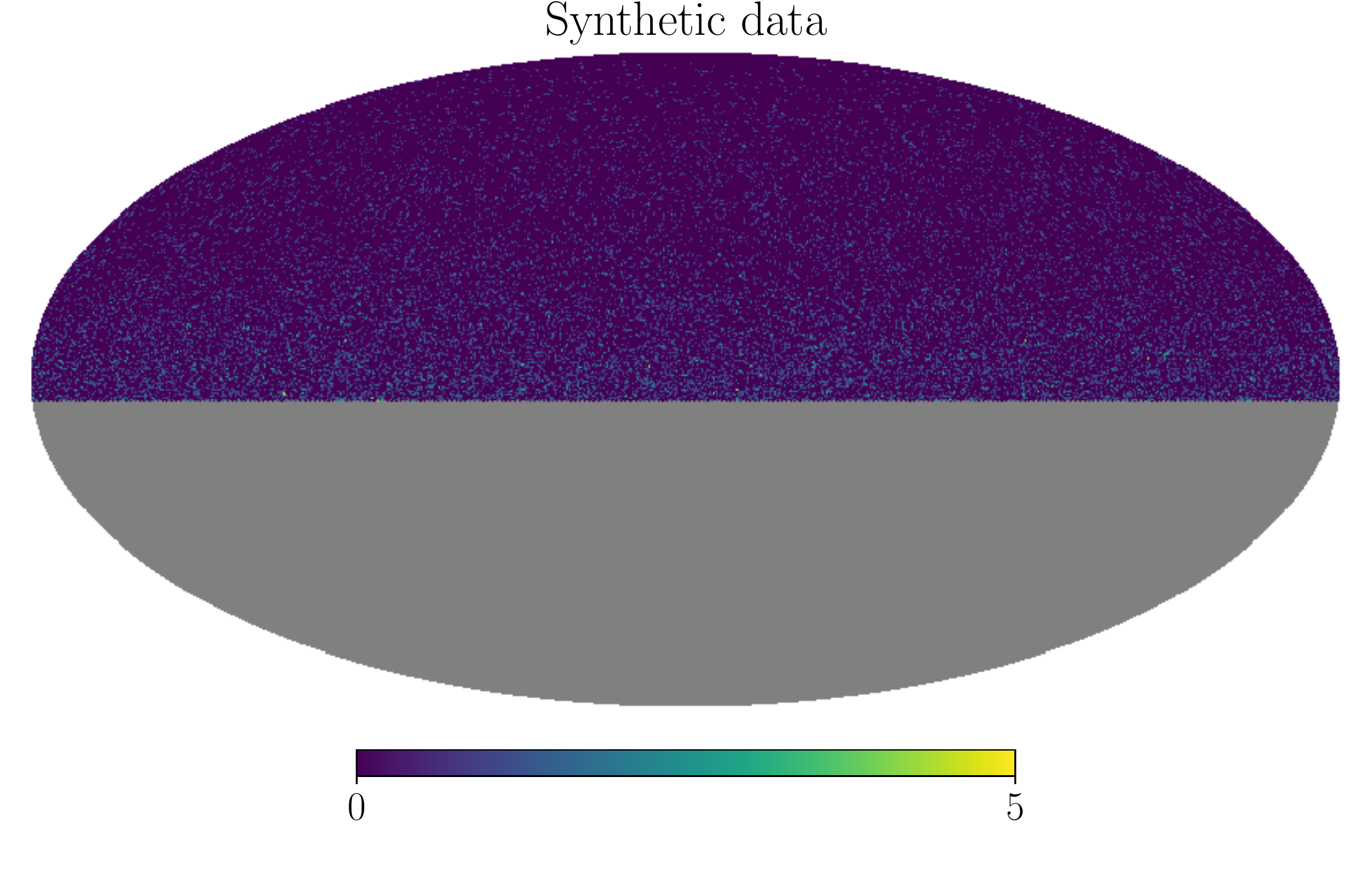}  
\caption{\label{fig:IceCubeCountsmap}
Number count maps of the IceCube all-sky three-year point-source public data (left) and synthetic data (right) between $10^4$ and $10^5$~GeV. The plots use HEALPix maps with NSIDE~=~128. The grey regions are masked (Dec $< -5^\circ$) to avoid muon cosmic-ray background. 
}
\end{figure}

\begin{figure}
\includegraphics[width= 0.5 \linewidth] {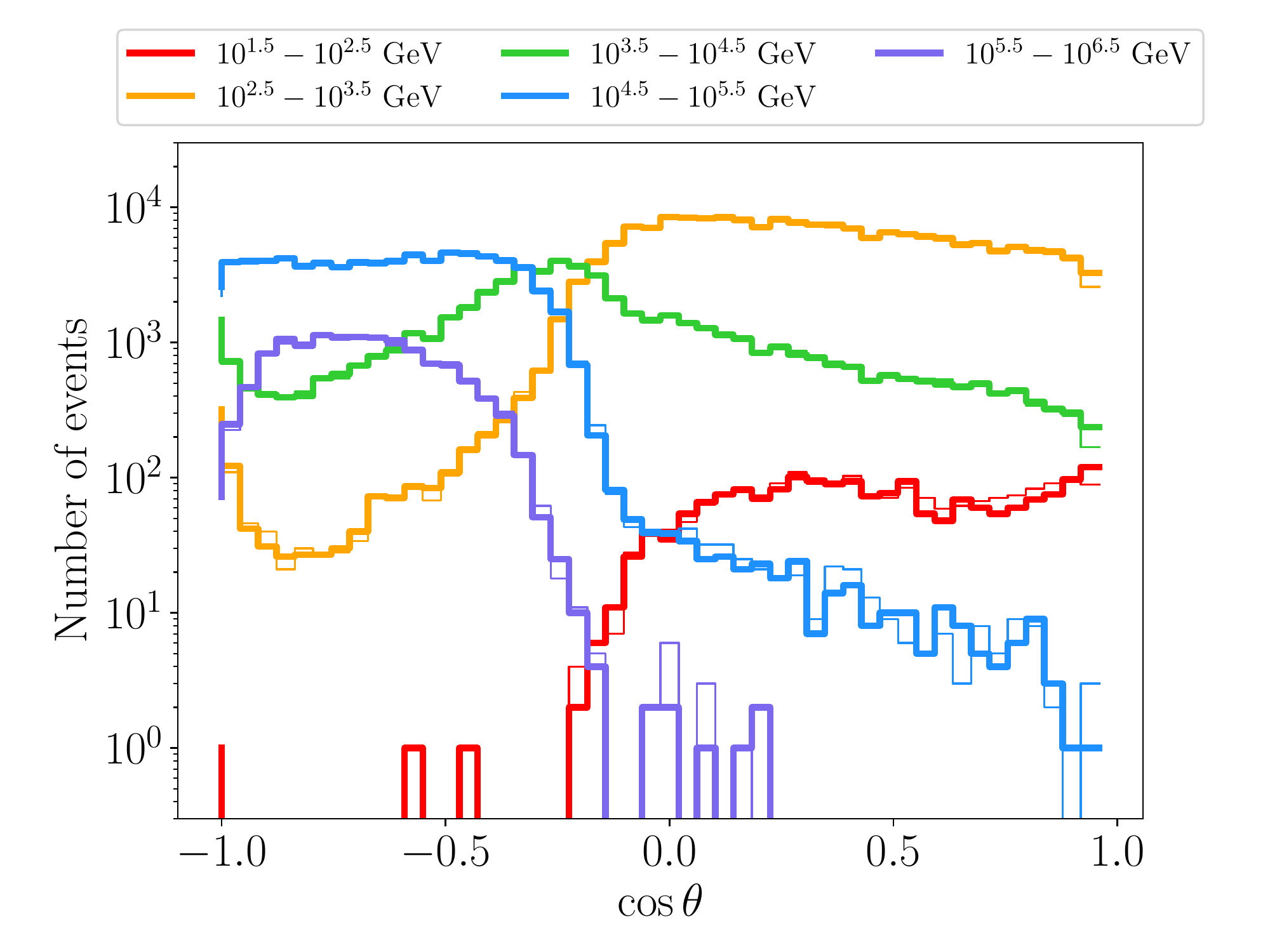}  
\includegraphics[width= 0.5 \linewidth] {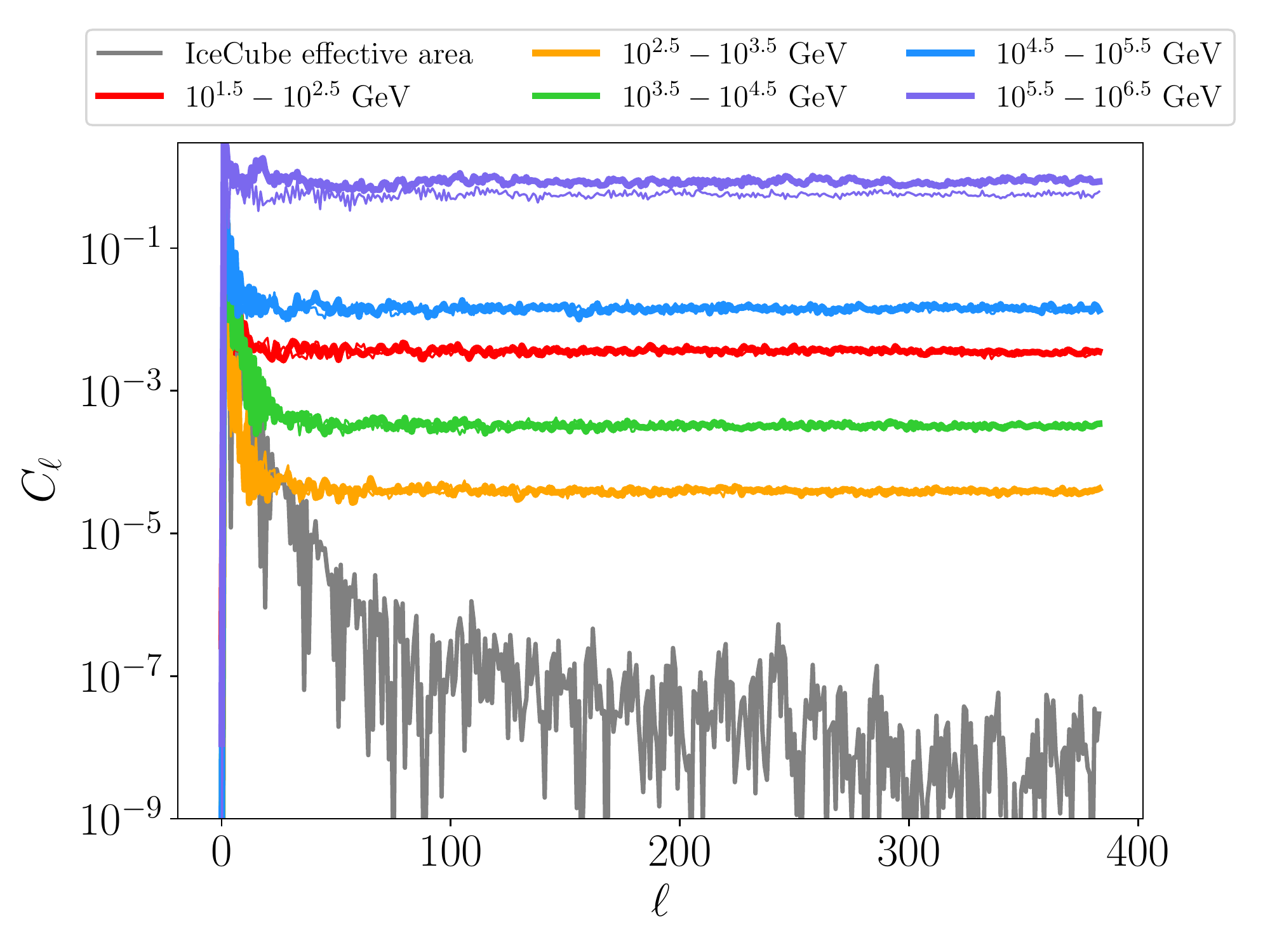}  
\caption{\label{fig:IceCubeAuto}
Left: histogram of $\cos\theta$ distribution of event counts in the IceCube point-source public data (thick curves) and synthetic three-year data (thin curves). $\theta$ is defined as $\theta = \pi / 2-{\rm Dec}$. The events are grouped into decade-wide energy bins indicated by the colors. Right:  power spectra of the over-densities of the events. For comparison, the power spectrum of the overdensity of a weighted effective area of IceCube is shown as a grey curve. The weighted effective area is computed by averaging the effective area of IceCube between $10^4$ and $10^5$~GeV assuming events follow an $E^{-3.7}$ energy spectrum. The $C_\ell$ spectra do not depend strongly on the energy binning or spectral weighting.
}
\end{figure}

The 2011-2012 data were taken with a full configuration of IceCube with 86 strings, while the 2010 data was taken with 79 strings. Each year of data contains about $10^5$ neutrino candidate events that passed the selection criteria described in \citet{Aartsen_2017}. In the northern sky, which is defined by \citet{Aartsen_2017} as $\theta > 85^\circ$, the sample is mostly composed of neutrinos since up-going muons are shielded by Earth. It is dominated by atmospheric neutrinos that follow a soft energy spectrum $dN/dE\propto E^{-3.7}$ \citep{PhysRevD.75.043006, Aartsen2015}. In the southern sky, the public data is dominated by cosmic-ray muons with energies up to 10~PeV. 

In each energy bin, we store the events in healpy maps with NSIDE~$=128$ in Celestial coordinates. The angular coordinates of a point on the sphere $(\theta, \phi)$ are converted to the right ascension (RA) and declination (Dec) by $\theta = \pi / 2 - \rm Dec$, $\phi =\,\rm RA$. 

The counts map of the IceCube point-source data in the energy range $10^4-10^5$~GeV is shown in the left panel of Figure~\ref{fig:IceCubeCountsmap}. 
The $\cos\theta$ distributions of the IceCube data in different energies are presented by the thick curves in the left panel of Figure~\ref{fig:IceCubeAuto}. The first two energy bins are mostly composed of atmospheric neutrinos in the northern sky.  The effective area in [$10^2, 10^3$]~GeV is much smaller than that in [$10^3, 10^4$]~GeV, and the first energy bin contains fewer events despite a greater atmospheric neutrino flux. 
 Above 10~TeV, muons from the southern sky dominate the distribution. The three-year data does not have events above 10~PeV. 

The auto-correlations of the over-densities of counts maps  are shown in the right panel of Figure~\ref{fig:IceCubeAuto}. The difference of the northern and southern skies, and declination-dependent differences in the same hemisphere lead to features at $\ell \lesssim 20$. The power spectra at $\ell\gtrsim 100$ are consistent with shot noise from Poisson statistics.

As the effective area of IceCube can vary notably across one decade of energy, we define a weighted effective area for a wide energy bin $i$ based on the effective area of finer bins $j$: 
\begin{equation}\label{eqn:weightedAeff}
A_{\rm eff, i} (\cos \theta) =  \frac{\sum_j \,A_{\rm eff, j}(\cos \theta)\,\left(E_{j, R}^{1-\alpha} - E_{j, L}^{1-\alpha}\right)}{\left( E_{i, R}^{1-\alpha} - E_{i, L}^{1-\alpha} \right)}\,,
\end{equation}
where $\alpha$ is the energy spectral index of events, and $E_{j,R}$ and $E_{j, L}$ are the energy at the right and left bounds of bin $j$. 
For comparison, the grey curve in the right panel of Figure~\ref{fig:IceCubeAuto} shows the power spectrum of the overdensity of the weighted effective area for neutrinos in the energy range $10^4-10^5$~GeV with $dN/dE\propto E^{-3.7}$. Above $\ell \sim 50$, the effective area has a much lower $C_\ell$ than the neutrino data and the galaxy sample (see Figure~\ref{fig:galaxy}). 
The smooth angular dependence of the effective area is unlikely to introduce features to the cross correlation of neutrino counts and galaxy maps.

\subsection{Generation of Synthetic Data}\label{appendix:syntheticData}
To generate synthetic atmospheric neutrinos and muons, we sample the zenith angles following the $\cos\theta$ distribution of the public data as shown in Figure~\ref{fig:IceCubeAuto} and assign random azimuthal angles to each event. The total event count in each energy bin is assumed to follow the Poisson distribution with a mean equal to the observed event count, $N_{\rm tot, i}$. 

To generate synthetic data that contain astrophysical neutrinos, we first assume $f_g=1$, and set the purity of astrophysical events in each energy bin, defined as 
\begin{equation}
f_{\rm astro, i}\equiv \frac{N_{\rm astro, i}}{N_{\rm astro, i}+ N_{\rm atm, i}}\,.
\end{equation}
To be consistent with the muon neutrino population in the point-source public data, 
we adopt $(f_{\rm astro, 1}, f_{\rm astro, 2}, f_{\rm astro, 3}) = (2.2\times 10^{-3}, 1.2\times10^{-2}, 0.15)$, which is based on the data and best-fit expectation from Monte Carlo simulation in the IceCube ten-year diffuse $\nu_\mu$ analysis (Figure~3 of \citealp{2019ICRC...36.1017S}).
Since we cannot differentiate the large muon population from neutrino events using the public data, we consider only the northern hemisphere for the astrophysical event generation and cross correlation study in this work. 
The synthetic data is thus composed of $N_{\rm atm, i}= N_{\rm tot, i}\,(1-f_{\rm astro, i})$ atmospheric events, which are generated the same way as pure atmospheric data, and $N_{\rm astro, i} = N_{\rm tot, i} f_{\rm astro, i}$ astrophysical events, as described below. 

Assuming that astrophysical neutrino sources share the same sample variance as the galaxies, the probability of an astrophysical event from pixel $j$ is given by the density of the galaxies $\rho_j$, with $\sum_j \rho_j = 1$. 
To simulate detection of these astrophysical events, we consider each energy bin i separately.   For bin i, we evaluate the ratio of the effective area in each direction j to the overall maximum effective area.  We treat these relative probabilities of detection $p_{i,j}$ as absolute probabilities and generate a total of $N_{\rm astro, i}^{\rm inj} = N_{\rm astro, i} / (\sum_j p_{i, j}\,\rho_j$) so that $N_{\rm astro,i}$ end up being detected.

The count map of a synthetic atmospheric event sample is compared with the actual IceCube data in Figure~\ref{fig:IceCubeCountsmap}. The zenith distributions and power spectra are shown as thin curves in Figure~\ref{fig:IceCubeAuto}. 

\section{Galaxy Catalogs} \label{appendix:GalaxySample}
\begin{figure}
\includegraphics[width= 0.5 \linewidth] {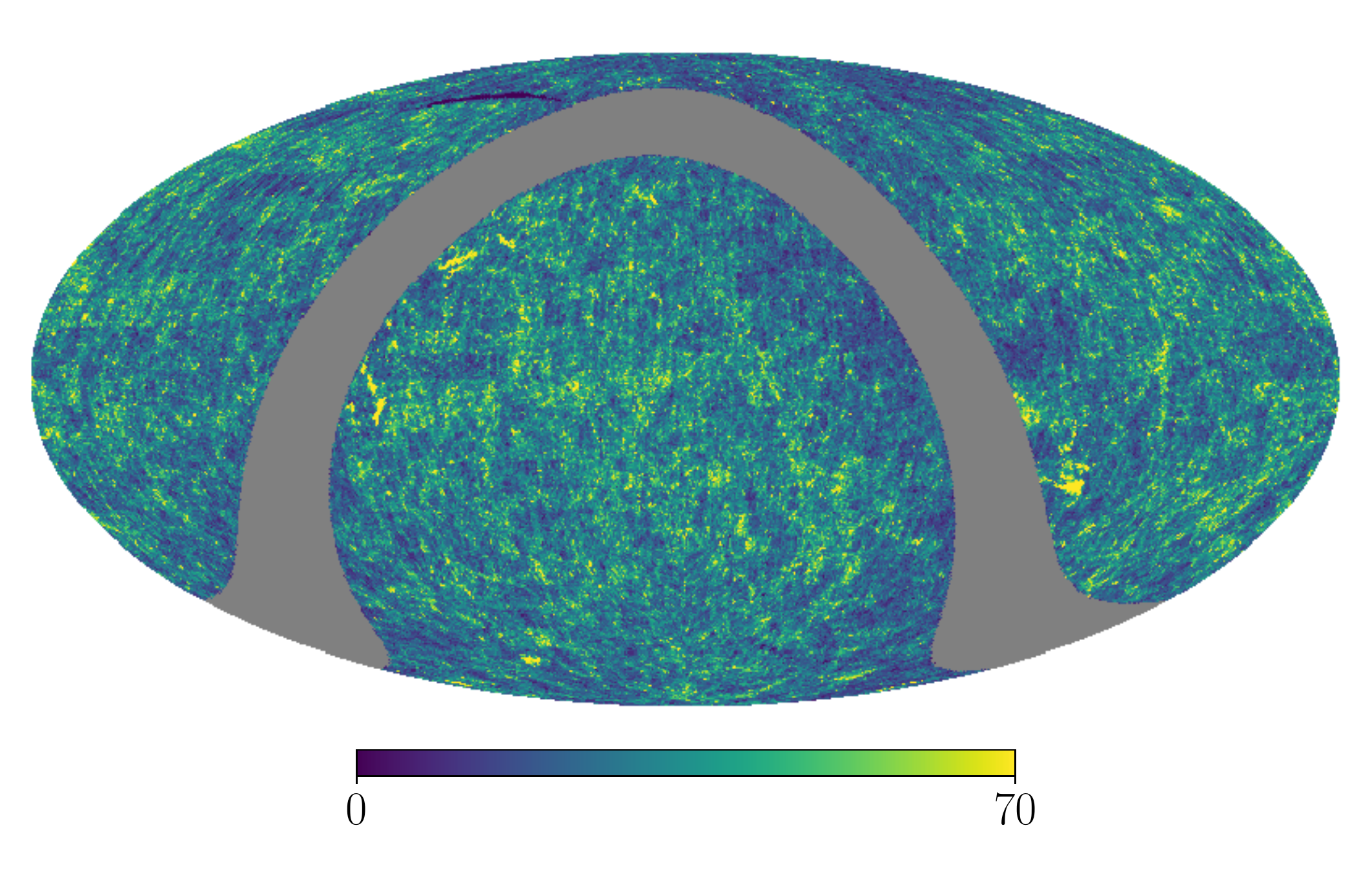}  
\includegraphics[width= 0.5 \linewidth] {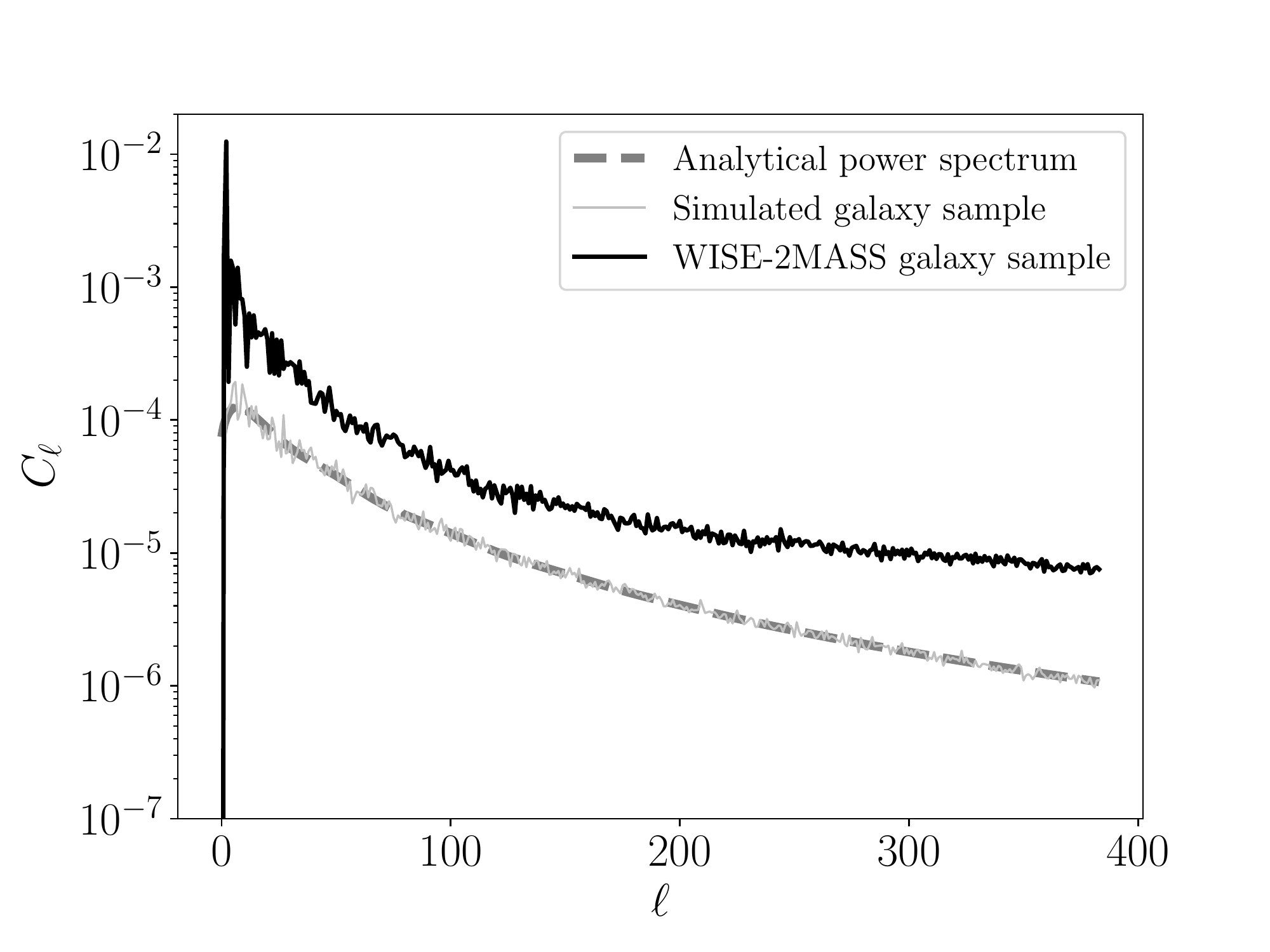}  
\caption{\label{fig:galaxy}
Left: number count map of the WISE-2MASS galaxy sample used in this work, constructed using  the Wide-Field Infrared Survey Explorer (\citealp{2010AJ....140.1868W}, WISE) and the 2-Micron All-Sky Survey (\citealp{2006AJ....131.1163S}, 2MASS) infrared databases following  \citet{2015MNRAS.448.1305K}. The grey region denotes a mask of the Galactic plane with $|b|<10^{\circ}$. The plot uses NSIDE~=~$128$. Right: power spectra of the WISE-2MASS galaxy sample (black curve) and a synthetic galaxy sample based on matter fluctuations computed analytically (see Appendix~\ref{appendix:analyticalGalaxySample}.)
}
\end{figure}
\subsection{Analytical Power Spectrum}\label{appendix:analyticalGalaxySample}

In order to generate an example of the expected $C_l$ of a sample of galaxies, we use the CLASS Boltzmann solver package\footnote{{https://lesgourg.github.io/class\_public/class.html}}. We choose the cosmological parameters to be the ones from the Planck 2018 best fit cosmology \citep{2018arXiv180706209P}. We further assume that the sample of galaxies has a constant comoving number density, within a redshift range of $0.2<z<0.6$. Finally, we assume that the galaxy sample has a bias parameter of $1.2$ with respect to the underlying matter fluctuations, and this value does not change as a function of redshift in the redshift range of interest. 
The analytical power spectrum is shown as a grey dashed curve in the right panel of Figure~\ref{fig:galaxy}. A synthetic full-sky galaxy sample is drawn from the analytical power spectrum and used for testing of the method.

\subsection{WISE-2MASS All-sky Infrared Galaxy Catalog}

We combined the 2MASS color data with the WISE photometry data to improve efficiency of star-galaxy separation.  We downloaded $\sim5$ million WISE-2MASS objects from the IRSA website \footnote{http://wise2.ipac.caltech.edu/docs/release/allsky/} that satisfied the selection criteria suggested by \citet{2015MNRAS.448.1305K}, and masked the region with Galactic latitude  $|b|<10^\circ$.  The resulting galaxy sample is expected to have  $<2\%$ stellar contamination and $>70\%$ galaxy completeness.  The left panel of Figure~\ref{fig:galaxy} presents the counts map of the WISE-2MASS galaxy sample.  The auto-correlation as computed with NSIDE$=$~1024, is shown as the black curve in the right panel of the figure.

\section{Projected ten-year results}
\begin{figure}
\includegraphics[width=0.43 \linewidth] {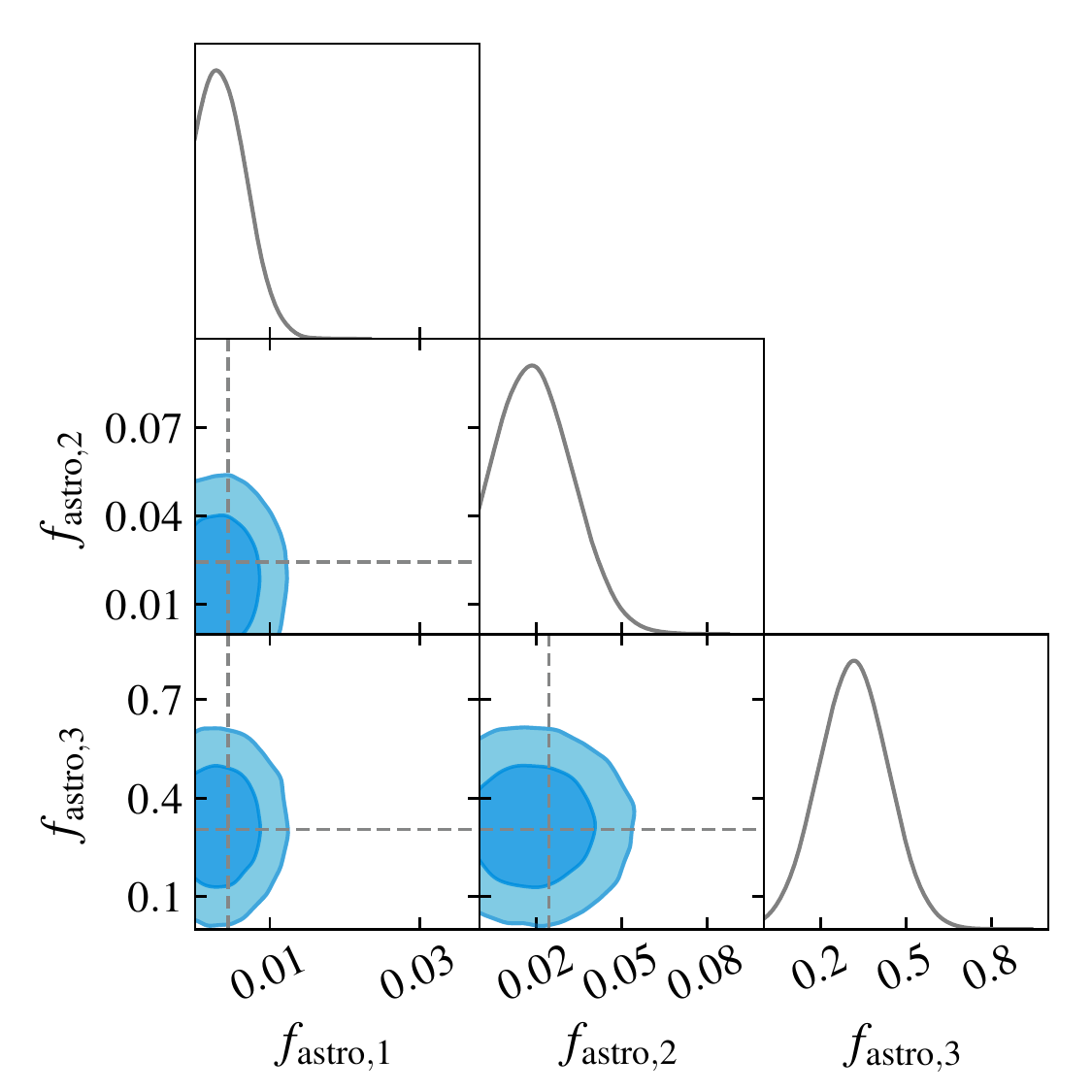} 
\includegraphics[width=0.57 \linewidth] {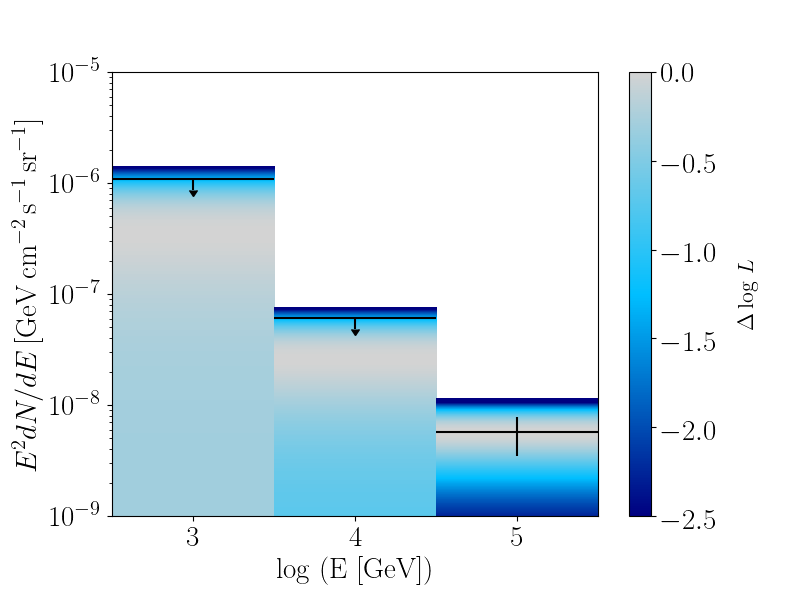}  
\caption{\label{fig:synthetic10yr}
Same as Figure~\ref{fig:IC3yr} but for synthetic ten-year data. Left: posterior distributions of $\mathbf{f}_{\rm astro}$ found by a MCMC sampling of the parameter space. The blue lines mark the input values of $\mathbf{f}_{\rm astro}$. Right: the best-fit energy spectrum of astrophysical neutrinos that follow the galaxy sample used for the analysis.    
}
\end{figure}

We present the projected results using synthetic ten-year point-source data in Figure~\ref{fig:synthetic10yr}. The synthetic data are generated using the cosine zenith distribution and effective area of the IceCube point-source data in 2012.  Compared to the three-year data, the ten-year data would contain more astrophysical events and better constrain $\mathbf{f}_{\rm astro}$. The median TS from $10^4$ realizations of synthetic data is $\sim 15$ in the optimistic scenario.
 

\end{document}